# Features that Predict the Acceptability of Java and JavaScript Answers on Stack Overflow

Osayande P. Omondiagbe
Landcare Research
Lincoln, New Zealand
omondiagbep@landcareresearch.co.nz

Sherlock A. Licorish
Department of Information Science
University of Otago
Dunedin, New Zealand
sherlock.licorish@otago.ac.nz

Stephen G. MacDonell
Department of Information Science
University of Otago
Dunedin, New Zealand
stephen.macdonell@otago.ac.nz

## Abstract

***Context***: *Stack Overflow is a popular community question and answer portal used by practitioners to solve problems during software development. Developers can focus their attention on answers that have been accepted or where members have recorded high votes in judging good answers when searching for help. However, the latter mechanism (votes) can be unreliable, and there is currently no way to differentiate between an answer that is likely to be accepted and those that will not be accepted by looking at the answer's characteristics.* ***Objective:*** *In potentially providing a mechanism to identify acceptable answers, this study examines the features that distinguish an accepted answer from an unaccepted answer.* ***Methods:*** *We studied the Stack Overflow dataset by analyzing questions and answers for the two most popular tags (Java and JavaScript). Our dataset comprised 249,588 posts drawn from 2014-2016. We use random forest and neural network models to predict accepted answers, and study the features with the highest predictive power in those two models.* ***Results:*** *Our findings reveal that the length of code in answers, reputation of users, similarity of the text between questions and answers, and the time lag between questions and answers have the highest predictive power for differentiating accepted and unaccepted answers.* ***Conclusion:*** *Tools may leverage these findings in supporting developers and reducing the effort they must dedicate to searching for suitable answers on Stack Overflow.*

## 1. INTRODUCTION

Stack Overflow is a popular question and answer portal used regularly by software practitioners exploring solutions to programming- and technology-related challenges faced during software development. Recent studies have shown that the majority of the questions that are asked on Stack Overflow receive one or more answers [1], and this forum is becoming a substitute for official programming languages' tutorials and guides [2]. Stack Overflow users (contributors) who post questions can select an answer, which is then typically regarded as the accepted answer. This allows the user who created the post to point to an answer that satisfies his/her question, and gives confidence to the user who wrote the answer. This can be convenient for other users browsing the platform seeking help for a similar issue, as they could locate an answer that may resolve their query in a short space of time, with confidence that the answer is acceptable to others. Answers may also be voted up or down depending on users' perception of their merit; however, this measure is said to be unreliable as users may manipulate the Stack Overflow voting scheme to enhance their reputation over others in the community [3]. In addition, while there is skepticism around the reliance of votes as a judgement of answers' suitability, answers take time to accumulate votes, and so, there are instances when votes are not available with which to offer preliminary judgement on answers.

Although answers on Stack Overflow may at times be accepted in 24 hours [3], in instances when there is delayed acceptance an automated mechanism may predict if an incoming answer will be acceptable (or accepted) based on specific features. Past studies have indicated that certain non-textual[1] factors affect the quality (and thus, acceptability) of answers retrieved in question and answer (Q&A) portals [4, 5]. Jeon, et al. [6] presented a systematic approach for extracting aspects of non-textual information that may be used to predict the quality of an answer in a Q&A pair. Other studies have also focused on using textual[2] features in finding high quality content in a Q&A platform [7]. This latter approach tends to overlook important non-textual features, while the former ignores those that are textual in nature. Therein exists the opportunity to evaluate features more comprehensively. In this paper, we examine the full range of features (textual and non-textual) that predict answers' acceptability. Our main assumptions are that: the best answer is the answer

[1] Relating to non-text (e.g., numbers, dates).

[2] Relating to text (e.g., words, phrases, sentences).

selected by the user who posed the question, and the best answer has the highest quality among the list of answers to a post. To the best of our knowledge, previous work did not examine the range of features that are common to Stack Overflow accepted answers, and how these may be used to differentiate acceptable answers. We thus set out to answer the following research question to guide our investigation:

***RQ**. Which features are most significant in distinguishing an accepted Stack Overflow answer?*

We believe that our outcomes could be of practical significance to the many practitioners who use Stack Overflow to answer questions and overcome challenges, as we provide understandings for the specific *important* attributes to look for when reading through potentially many Stack Overflow answers.

The remaining sections of the paper are organized as follows. Section 2 examines how other researchers have studied the quality and acceptability of answers in Q&A settings. Section 3 explores the structure of Stack Overflow processes. In Section 4, we present the research methodology, detailing our data extraction, data sampling and procedures, text data processing, features selection and modelling and analyses. Section 5 reports our results in answering the research question above. We next discuss our findings and explore implications in Section 6, before considering threats to the work in Section 7. This paper then concludes in Section 8, where we provide a summary of our work and recommendations for future research.

## 2. BACKGROUND

While votes accumulated could influence answer acceptability [8], this measure is unreliable as users may manipulate voting schemes to enhance their reputation [8]. We therefore investigate the features that are most significant in predicting answer acceptability, and review works that have studied the attributes that influence the quality and acceptability of answers in Q&A settings. For instance, Jeon, et al. [6] focused on non-textual features when examining Q&A forums and proposed a framework to predict the quality and acceptability of content in a collection of Q&A pairs using clustering and maximal entropy. They extracted a range of non-textual features, and the length of the answer was utilized in determining answer acceptability. Larkey [9] used the length of answers to estimate the quality and acceptability of online writing. This feature was also supported by others (e.g.[10]), who observed that the quality of answers correlated with their length. [5] went on to demonstrate a graphical based approach to determining high quality content in a Yahoo Q&A dataset. They focused only on non-textual features, and found that answer length was the dominating feature among several other features in determining the quality of answers leading to their acceptance. This stream of work shows that non-textual features can be significant predictors of answer quality and acceptance in some contexts, and especially the number of words that are expressed in contributors' responses.

Other works have examined non-textual features with a slightly different focus. For instance, Burel, et al. [11] combined user and thread features to predict the best answer by using logistic regression. They found, contrary to the outcomes of previous work, that answer length was not correlated with the best answers that were provided by contributors to Q&A portals. In fact, in examining an online community website where users with no prior knowledge interacted with each other, users' reputation was shown to be the strongest predictor of the quality of posts [12]. In order to compute users' (contributors') reputation in a Q&A setting, other work has combined both social network analysis (SNA) metrics and user rating [13]. The outcomes here show that no one variable may predict answer quality and acceptance, and in fact, isolated predictors may be evident by chance.

The need to increase the range of features used for training models has thus led other researchers to combine both textual and non-textual features. Combining textual and non-textual features was seen in the work of Blooma, et al. [14]. They predicted the best answer in a stack of answers by using a Bayesian model, and concluded that the best answer was greatly influenced by textual features. Other researchers have also combined both forms of features to extract high quality content in online forums. For instance Jizhou, et al. [15] combined structural information with textual and non-textual features to extract high-quality pairs (discussion threads) from an online discussion forum. Buse and Weimer [16] explored users' metadata and other textual and non-textual features as input to a support vector machine to determine the best answer in Yahoo Q&A datasets. Both groups of authors found that the user profile variable was a strong predictor of content quality. These findings underscore the need to investigate both textual and non-textual features for their predictive power when exploring Q&A forums.

Understanding the strength of various predictors could go some way towards identifying specific posts in software development Q&A forums, where practitioners are seen to rely heavily on such content for problem solving [2]. Stack Overflow, in particular, has become a central portal for developers' support [1], and has attracted a large amount of research effort [17]. While works have attempted to predict accepted answers on this portal (e.g., via votes [18] and comments [19]), there has been less effort aimed at studying the full range of features towards predicting answers acceptability on Stack Overflow. We thus investigate this issue.

## 3. STACK OVERFLOW

Stack Overflow was built around four design decision mechanisms[3] to ensure maintenance of content quality. These are the **voting** mechanism[4], where users can up-vote or down-vote answers they like or dislike respectively. The **Tag** mechanism is used to organize Stack Overflow questions into groups; users are allowed the option to assign at least one (or up to five) tag(s) to their question. Users apply the **editing** mechanism to edit and refine their Q&A over time, thereby providing more reliable and precise information to the community. Finally, **badges**[5] are given to contributors once their contributions have reached a certain threshold to reward them for their effort (as a form of prestige).

[3] http://www.stackoverflow.com

[4] https://stackoverflow.com/help/why-vote

Tags are particularly noteworthy for Stack Overflow users. As noted above, the user that registers the question on Stack Overflow is also allowed to label the question by placing a tag. The tag indicates the group(s) the question belongs to; for example, "computer programming", "API" or "library". A more granular classification may also be provided, e.g., PHP, MySQL, C#, and so on. Each tag is presumed to be representative of the question, and the Stack Overflow system automatically suggests tags for users at the time of entering questions. Users of Stack Overflow sometimes use the tag mechanism to search for Q&A pairs, and the need for an accurate tagging system has led some researchers to develop models that predict the tag for given questions [20].

Other mechanisms have been provided for encouraging contributions on Stack Overflow. As noted above, in order to encourage users' activities the concept of badges is used. Badges are usually earned by being helpful and performing particular tasks on Stack Overflow, which leads to the user building a positive reputation. Bosu, et al. [8] in their study of members' reputation via badges in Stack Overflow found that a new user would usually gain status quickly and move on to a higher position within the Stack Overflow community by gaining badges. Finally, as mentioned earlier, users of the Stack Overflow website are encouraged to use the editing mechanism to edit and refine their Q&A over time. This feature is seen to be positive in terms of encouraging progressively more acceptable questions and answers over time. The abovementioned practices make Stack Overflow an interesting platform to study. We examine the specific research methodology used in this work next.

# 4. RESEARCH METHODOLOGY

In this section, we present our data extraction process (Section 4.1), efforts towards data sampling and preprocessing (Sections 4.2 and 4.3), feature selection (Section 4.4) and modelling and analyses (Section 4.5).

## 4.1 Data Extraction

The Stack Overflow data dump is published by Stack Exchange in XML format[6], and is available for researchers to perform various forms of analyses. We used the dataset added to the archive on September 12, 2016, which is divided into several XML files, consisting of posts, comments, tags, badges, post history, post links, users and votes. The Post XML file contains the questions and answers, and also the tag variable which is linked to the Tag XML file. The Tag XML file has the name of tags and associated IDs for the names. The PostLinks XML file provides linkages to related Post IDs for the various types of posts (question or answer). The Comment XML file has comment text and corresponding Post ID (the post for the associated comment(s)), thus linking back to the Post XML file. A similar linkage exists for the Votes and Badges XML files, which are linked to the Post XML and Users XML files respectively. The Users XML file contains user (contributor) data, covering contributors to both questions and answers. The combined dataset extracted was 90 gigabytes, and it contained over 35 Million posts (questions and answers). In order to simplify our working processes we created a Python script to convert all XML files to plain text and dumped the content into their corresponding tables in a SQLite database (e.g., the Posts.XML file was dumped into Post table). We were not interested in analyzing the entire dataset because our objective was to understand the features that make answers acceptable, and to accomplish this we sampled records that were useful for our study, i.e., posts with more than one answer, and which also have an accepted answer.

## 4.2 Data Sampling and Procedures

To answer our research question, we needed to understand the features that make answers acceptable; thus, we extracted questions that have accepted answers only. We studied the top 50 tags in our dataset to gain insight into the numbers of questions and answers that have been contributed to each tag and found that JavaScript, Java and C# were the three top tags. Since we are interested in understanding the strength of the various features that make an answer acceptable we reduced the scope of our dataset by including only questions with two or more answers. These questions were also required to have a corresponding accepted answer. To reduce our dataset further, to optimize our algorithms' execution, we selected records for the top two tags only (Java and JavaScript), in light of these records having proportionally more answers than others. We explored the dataset associated with those two tags further to examine the trend of questions and answers over time, finding that the final 3 years (2014, 2015 and 2016) have recorded the highest number of questions (and associated answers) on Stack Overflow. While questions may be forwarded around the scale of data we analyzed (refer to Section 7 for additional details), our dataset comprised nearly 250,000 records, and we know that Stack Overflow data are generalizable across time and languages [21].

In this paper we are only interested in the predictive power of the features of accepted answers, and so we eliminated all features that do not relate to what makes an answer acceptable, prior to our analyses. We discarded the badges table because badges are given based on how many questions and answers a contributor provides; and do not relate to what the community thinks about those questions and answers. In addition, badges differ from the reputation score[7] of contributors, with the latter attribute being more suitable for understanding a contributor's worth to the community. We also discarded the post history, post link and user tables as these tables were not held to contain information related to our research question. Furthermore, we discarded answers that were given and accepted by the same user who posted the question, and answers that were not linked to any registered user.

[5] https://stackoverflow.com/help/badges

[6] https://archive.org/details/stackexchange

[7] Based on the quality of the question and answer the user has posted.

Our dataset for experimentation thus consisted of 249,588 posts. In line with our sampling procedure above, these posts comprised questions from 2014, 2015 and 2016 that have at least two answers, of which there is one accepted answer. We had a slightly unbalanced dataset (in terms of the number of accepted and unaccepted answers), where the number of accepted answers was 88,607, compared to 160,981 answers not labelled as accepted. We randomly selected 70% (174,711) of the data for training, and the remaining 30% (74,877) was used for testing when performing our predictive modelling (refer to Section 4.5).

### 4.3 Text Data Preprocessing

In this section we present the approaches used for preprocessing our text data. We found that Stack Overflow Q&As are usually stored in Markdown (HTML), and code blocks are always placed between a tag block called "code". An example of an answer with a code block is shown below:

```
"<p>I do not think it is necessary and I would ...  2014 </p>
 <code> def plot: a=c[1,2,3] </code>"
```

Our first step was to extract these code blocks by separating them from the text body, before exploring both the code and text parts of answers in detail. Preprocessing of text documents is a vital task during text mining. This is because retrieving meaningful information from preprocessed texts is easier when compared to natural language because text documents are usually represented as a bag of words with various dimensions [22]. These dimensions can be reduced by applying preprocessing techniques such as stop word removal and tokenization [23]. The subsections below detail our text data preprocessing steps, which consist of stop word removal, stemming and tokenization.

***(1) Elimination of Stop Words:*** When retrieving information in a text document, many words do not add meaning to the sentences in which they belong [23] . These words are usually grouped as the most common words used in English language, and are classified as prepositions, conjunctions or articles [23]  . They are mostly used to join words in a sentence. Examples of stop words include: "above", "but", "an", "anything". These words are irrelevant because they do not provide any useful information during information retrieval. By eliminating stop words from a document containing text, we are also reducing the size of the document index structure, which ultimately results in improved performance of our text mining algorithms. We used the nltk corpus[8] library to remove stop words.

***(2) Stemming:*** Stemming involves reducing derived words from a corpus into their root form; e.g., "coming" to "come". This is done because the related word will map back to the root form giving the same meaning [23]. A stemming algorithm removes prefixes and suffixes and produces a stem. Stemming is said to be an important step in text mining because this exercise reduces variation of words which have the same root form from those that have a common meaning [23]  . The stemming process also further reduces the size of the document index structure. We applied the popular Porter stemming algorithm [24] to our dataset.

***(3) Tokenization:*** After performing stemming and stop word removal, the texts in each record were separated into individual words by removing punctuation, whitespace and alphanumeric characters. This process is called tokenization. The tokenized words are separated by spaces in each record and are passed as input to our text mining algorithm. The main aim of tokenizing the sentence is to identify meaningful keywords in each sentence [25].

### 4.4 Features Selection

In this section we describe the features used in our models and the reason for selecting those features. We grouped our features into four categories (code, textual, non-textual and user features), and describe these in the following subsections.

***(1) Code Features:*** Code features describe the properties of the code found in each answer. Previous work on code readability concluded that lines of code and the average number of identifiers (e.g., constant and parameters) per line predict code readability [14]. Based on this finding, we extracted the number of lines present in the code (**NumberOfcodeLine**) and number of identifiers (**Codelength**) as features in our models.

***(2) Textual Features:*** Textual features describe the properties of each text answer. In order to extract meaningful textual features we examined the work of Blooma, et al. [14]. Their work pointed out that accurate answers may have syntactic relations to associated questions. Thus, we computed the textual similarity between questions and answers. The textual similarity shows how close the question and answer are to each other, allowing us to map word usage across these pairs. Beyond this feature, we computed the polarity of answers (i.e., measurement of emotional content), vector concordance similarity between the questions and answers, length of answers (in words), and the number of sentences in given answers. We now list the textual features extracted, the methods used for extraction, and justification for selecting these features where necessary:

Question and Answer Similarity (**TFAnswerText**): This is a similarity score evaluated based on the textual similarity of a question and answer pair. This feature was computed by first converting the pre-processed text (question and answer) to a vector matrix of term frequency-inverse document frequency[9] (tf-idf) features, and comparing the similarity of the question and answer vector by using cosine similarity defined in (1):

$$w_{t,d} = nf_{t,d} \times idf_t \qquad (1)$$

$$\text{where} \qquad idf_t = \log \frac{N}{df_t}$$

$$nf_{t,d} = 0.5 + \frac{0.5 \times tf_{t,d}}{max_{t\in} tf_{t,d}}$$

[8] http://www.nltk.org/howto/corpus.html

[9] A statistic which reflect how important a word is in a document.

Where:
$tf_{t,d}$ = number of times a term (t) appears in a given Q&A pair/document (d);

*Document frequency* ($df_t$) = number of Q&A pairs in which a term (t) appears;

*Inverse document frequency* ($idf_t$) = how much information a given term provide, where N represent the total number of Q&A pairs and $df_t$ is the number of Q&A pairs in which term t appears. It is expressed as the logarithmically scaled inverse fraction of the Q&A pairs (document) containing the term [26] .

The questions and answers were also converted to a vector matrix of term frequency-inverse document frequency vectors which was then used to compute the similarity between questions and answers using cosine similarity (using (2)):

$$\text{Similarity (Q, A)} = \frac{Q \cdot A}{|Q||A|} \quad (2)$$

Where:
|A| = magnitude of answer vector;

|Q| = magnitude of question vector;

Q. A = dot product of the question and answer vector.

We also compute the similarity between the question and code in answers using the same approach (**TFAnswerCode**).

Polarity (**TextPolarity**): This feature measures the emotional content of answers. The sentiment of answers could be positive, negative or neutral. This feature was extracted because it is assumed that a positive response will be more likely to be helpful than a negative response [27]. However, previous research has not studied the importance of such a feature in a Q&A setting. That said, beyond Licorish and MacDonell [28], other work has shown that specific software development tasks attract various forms of developers' emotions [29]. We use Textblob[10], a python library, to assign a polarity score to each of the answers. Textblob was chosen because it provides a simple API for performing most natural language processing (NLP) tasks; for instance, part-of-speech tagging [25] .

Vector Concordance Similarity (**TextualSimilarity**): We computed the count of every word that occurred in the question and answer separately. Then, we converted these to vectors, and calculated how similar the vectors were.

Given the work of Blooma, et al. [14], which measures the accuracy and completeness of an answer by looking at the numbers of sentences and words in an answer, and finding concise answers to be more accurate, we extracted the following three additional textual features:

Length of Answer (**NumberOfWord**): This is the number of words in each answer after stop words are removed.

Number of Sentences (**NumberOfSentence**): This is the number of sentences in each answer.

**UrlCount**: This is the number of URLs present in each answer. We decided to include this feature because URLs were embedded in some of the answers. An URL in an answer could mean the poster is trying to point the reader to more resource online, making this feature worthy of inclusion.

***(3) Non-Textual Features:*** Non-textual features show the non-textual properties related to each Q&A, comprising a feature-set that is structured (and not derived from text). We examined all the available attributes in our dataset to generate these features, comprising the following:

Response Time (**Timelag**): This is the difference in time between when a question was posted and when it was answered. This was calculated by subtracting the answer date from the question date, and the difference was then converted to milliseconds (for all answers).

Number of Comments (**CommentCount**): This feature was present in our initial dataset, and we decided to include it in keeping with earlier evidence, which established that good answers usually attract numerous comments [30].

**AnswerCount**: This feature was present in our dataset. It shows the number of answers a given question has. This measure was also used to subset our dataset, as we needed questions with more than one answer. We included this feature because it is anticipated that numerous answers may result if a question is interesting, which may influence answer quality.

AnswerScore (**Score**): This is the voting score for each answer, and was initially present in our dataset. Stack Overflow calculates this score by subtracting the number of downvotes[11] from the number of upvotes[12]. Notwithstanding some concerns around users' gamification of votes [8], we anticipated that this feature would influence how the community regard answers.

**ViewCount**: This feature shows how many people have viewed an answer. We included this feature because it was part of our initial dataset, and from the visualizations in Fig. 1 it is noted that there was not much difference between the view count of accepted and unaccepted answers. Here "Frequency Count" (y-axis) is the number of answers with various view counts (x-axis). This pattern could be evident because answers are usually viewed by Stack Overflow community members before they are accepted and scored.

***(4) User Features:*** User features describe the characteristics of the user writing the answer. We included the reputation of the user who posted the answer (**Reputation**) and how long the user has been a contributor to Stack Overflow (**SignUpDateTimeLag**) as part of our list of features. Previous studies have reported that individuals with a strong and established reputation usually write good answers (see for example: [8]). The features above were all used to distinguish acceptable answers, where we excluded all features which occurred after the user accepted an answer for a given question.

---

[10] http://textblob.readthedocs.io/en/dev/

[11] The number of people in the Stack Overflow community who think the answer was not helpful.

[12] The number of people in the Stack Overflow community who think the answer was helpful.

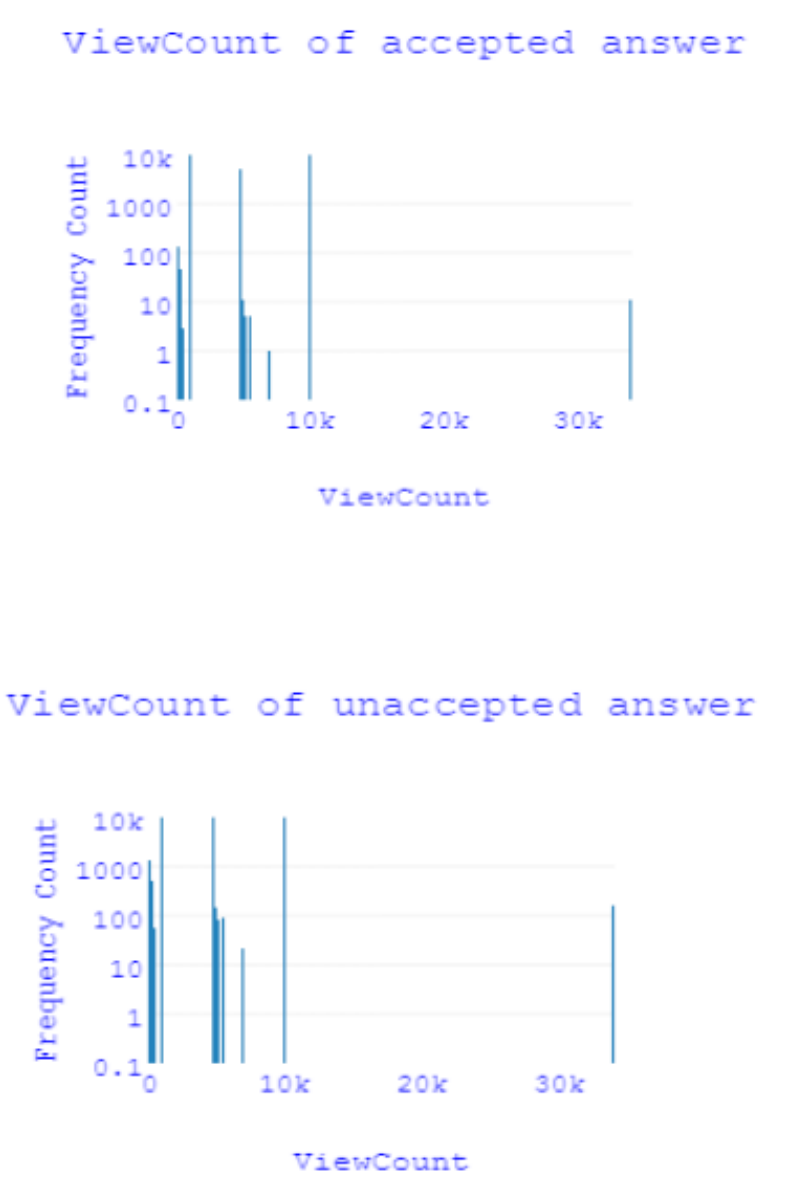

Fig 1. View counts for accepted and unaccepted answers

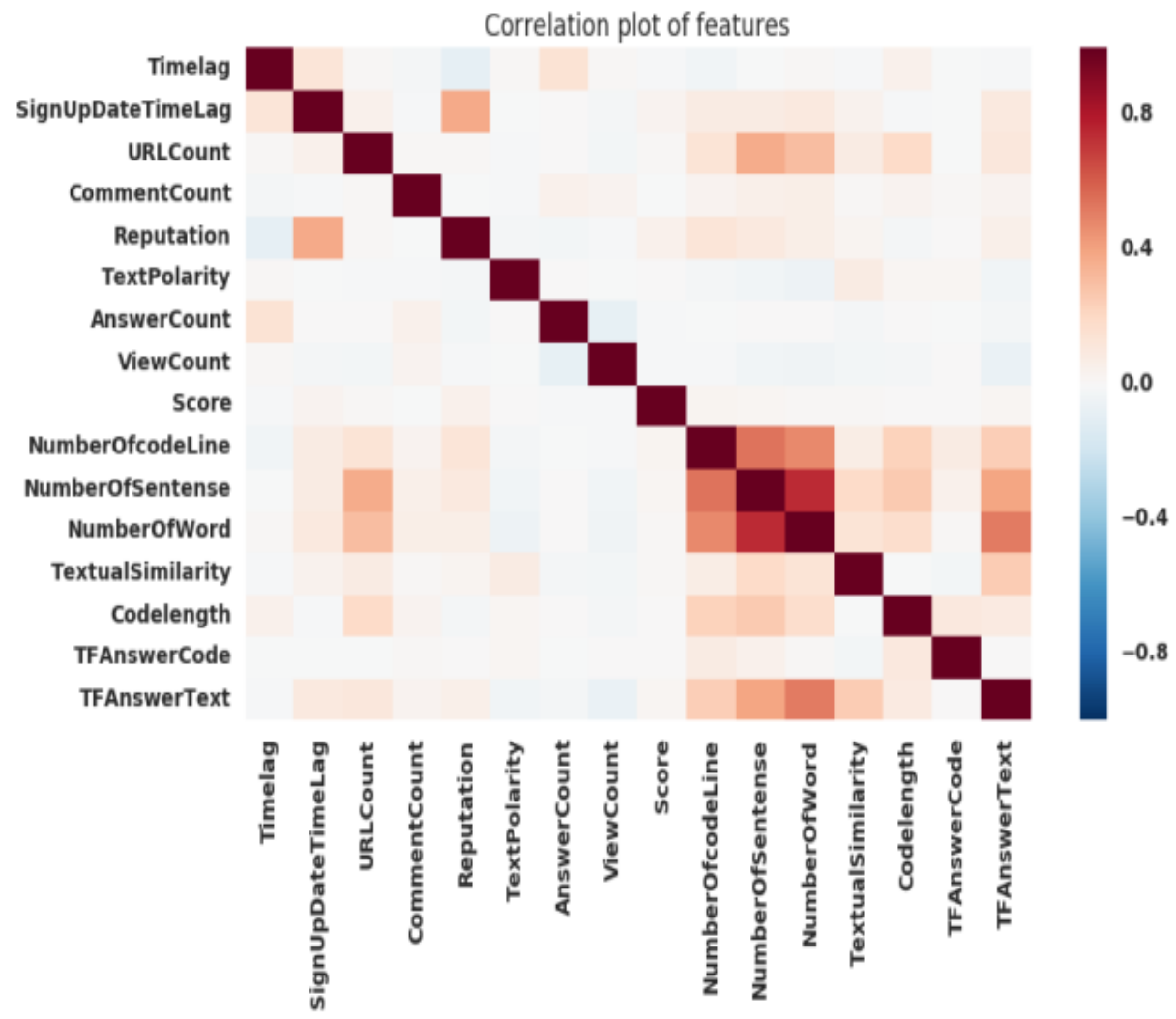

Fig 2. Correlation plot of features

### 4.5 Modelling and Analyses

We used a random forest to build a model that classifies an answer as either accepted or unaccepted. Given the exploratory nature of the study, the random forest method was chosen as it takes many input variables without the need for replacement [31]. In addition, random forest estimates the importance of each variable in the classifier, while using an out-of-bag[13] estimator to estimate the classification error when sampled with replacement [32]. This approach helps to prevents overfitting, and does not require the use of k-fold cross validation (which can be computationally intensive with numerous data points) [31].

Due to the unbalanced nature of our dataset, we employed the recommended synthetic minority over-sampling 'SMOTE' algorithm [33], and the adaptive synthetic sampling 'ADASYN' algorithm [34]. We created a parameter grid to aid sampling, where six hyperparameters were chosen: **n_estimators** (100 to 1200, stepping through by 100, which makes a total of 12 steps), **max_features** (sqrt of 14 =3.75, approximate 4), **max_depth** (10 to 110 (step through 11), which makes a total of 10 steps), **min_samples_split** (set to 3), **min_samples_leaf** (set to 3), and **bootstrap** (set to *True*, i.e., 1). Use of the six parameters selected above resulted in a total search space of 12*4*10*3*3*1=4,320. This may be assessed as computationally expensive, hence, we did a random search to sample a wide range of these parameters. We used 100 iterations and 4-fold cross validation to fit our model and retrieve the best parameters, including; bootstrap: True, max_depth: 60, max_features: auto, min_samples_leaf: 3, min_samples_split: 8, n_estimators: 200. We then applied the two sampling techniques (SMOTE and ADASYN) and the best parameters mentioned above to tune our model.

In providing triangulation for our random forest outcomes, we repeated our experiment with a more complex model – a neural network – to classify answers. This was chosen because such models are universal approximators, able to learn complex relationships manifested in multi-dimensional datasets [35]. Our neural network model had five hidden layers, and a stochastic gradient descent (SGD) was used as our optimizer because our data are not sparse, and SGD is faster and less prone to unfavorable local minima [36] . We also applied the two sampling algorithms (SMOTE and ADASYN) when modelling using our neural network. We seeded the random split function in order to have the same split for each execution of our algorithm (i.e., random forest and neural network). Modeling was done through the use of the Python scklearn library. We evaluate both models in the next section and provide other associated results.

Table 1. Information gain for all features

| Feature | Information Gain |
|---|---|
| Timelag | 0.873 |
| URLCount | 0.432 |
| CommentCount | 0.563 |
| Reputation | 0.893 |
| TextPolarity | 0.567 |
| AnswerCount | 0.445 |
| ViewCount | 0.563 |
| Score | 0.456 |
| NumberOfcodeLine | 0.612 |
| NumberOfSentence | 0.654 |
| TextualSimilarity | 0.534 |
| Codelength | 0.456 |
| TFAnswerCode | 0.579 |
| TFAnswerText | 0.467 |
| SingupDateTimeLag | 0.234 |
| NumberOfWords | 0.345 |

## 5. RESULTS

To avoid multi-collinearity, we executed a Pearson's correlation plot for all of the features extracted. This informed our modelling, where we selected only the feature pairs where the root mean square was < 0.7, in keeping with convention [37]. Fig. 2 shows the correlation matrix for all features, where darker squares reveal variable convergence ($r \sim 1$, i.e., associations were linear or close to linear). We also compute the mutual information gain for each feature [37], as presented in Table 1. Of note in Fig. 2 is that "number of words" had a strong correlation with "number of sentences" (coefficient = 0.82), and the former variable has lesser information gain in Table 1. Thus, "number of words" was discarded prior to modelling. In addition, "sign-up date" correlated with "reputation" (coefficient = 0.76), with the former variable also having lesser information gain, and hence, this variable was also removed prior to executing our final models. We included all other features with information gain > 0.4 [38] in Table 1.

[13] Method that works by estimating the error by leaving out a sample of the data.

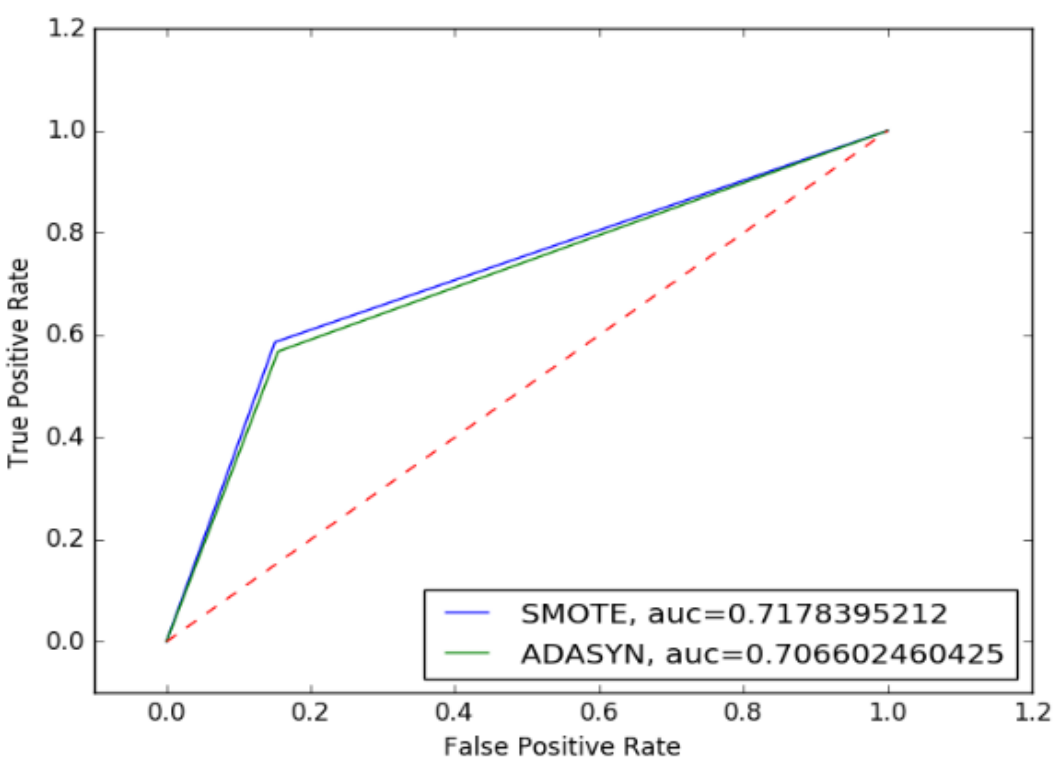

Fig 3. ROC curve of baseline model

Fig. 3 shows the ROC curves for our random forest model, depicting accuracy of 70.6% when sampling with the ADASYN algorithm, and 71.7% when sampling with the SMOTE algorithm. We observe a similar outcome for our neural network model (with accuracy of 70.9% when sampling with the SMOTE algorithm, and 69.8% when sampling with the ADASYN algorithm). In Fig. 3 the area between the blue and dashed red lines measures how useful our random forest model performs over a random guess when our dataset was sampled using the SMOTE algorithm. The green line measures how useful our model performs over a random guess when our dataset is sampled using the ADASYN algorithm. This illustrates how accurately the random forest model could separate accepted answers from unaccepted answers. In fact, our random forest (SMOTE sampling) accuracy of 71.7% is considered to be "fair" at separating accepted answers from unaccepted answers.

To understand which features predict answers' acceptability, we examined the features and their contributed weights in our random forest and neural network models. Table 2 shows the features and their corresponding weights in both models when using the SMOTE algorithm, where it is noted that all of our features had a positive direction (e.g., the longer the code provided in answers and higher the reputation of the contributor the greater the likelihood of acceptance). In Table 2 it is observed that the length of code (Codelength), time it takes to post an answer (Timelag), and reputation of the answerer (Reputation) had the largest coefficients in both random forest and neural network models (Codelength = 0.234 and 0.153, Timelag = 0.162 and 0.152, Reputation = 0.143 and 0.149, respectively). Table 2 shows that the textual similarity between a question and answer pair (TFAnswerText) was also noteworthy in predicting Stack Overflow acceptable answers, with coefficients of 0.124 and 0.133 returned for our random forest and neural network models respectively. Other features did not consistently predict answers' acceptability in both models, with their coefficients tending to be higher in the random forest model. For instance, the number of lines of code (NumberOfcodeLine) and text polarity (TextPolarity) features recorded coefficients of higher magnitudes for the random forest model in Table 2.

Overall, our outcomes here confirm that several features have a positive impact on answer acceptability, with these features accounting for 71.7% and 70.9% accuracy in the prediction of acceptable answers when sampling with the SMOTE algorithm, and 70.6 % and 69.8% when sampling with the ADASYN algorithm for our random forest and neural network models respectively.

We computed a confusion matrix for the random forest and neural network models for the different sampling techniques used and observed good to excellent precision and recall values. Precision and recall results for our SMOTE sampling technique were: Random Forest = 88.25, 73.29 and Neural Network = 87.29, 72.15, respectively. When sampling with the ADASYN technique our precision and recall results were: Random Forest = 85.04, 71.07 and Neural Network = 83.13, 69.45, respectively. Our Matthews Correlation Coefficients (MCC) for the respective models when sampling with the SMOTE technique were moderate (Random Forest = 0.39 and Neural Network = 0.34).

This outcome suggests that, on the whole, in the absence of an accepted answer, it is feasible to inform the software engineering community's selection of Stack Overflow answers that are acceptable (or are more likely to be accepted) based on specific features. This evidence may thus support developers' timely identification of suitable Stack Overflow answers. We discuss our outcomes and their implications in the following section.

Table 2. Coefficients for features of random forest and neural network models

| Feature | Random Forest Coefficient | Neural Network Coefficient |
|---|---|---|
| *Timelag* | 0.162 | 0.152 |
| URLCount | 0.044 | 0.021 |
| CommentCount | 0.043 | 0.008 |
| *Reputation* | 0.143 | 0.149 |
| TextPolarity | 0.065 | 0.007 |
| AnswerCount | 0.023 | 0.015 |
| ViewCount | 0.025 | 0.057 |
| Score | 0.052 | 0.047 |
| NumberOfcodeLine | 0.087 | 0.023 |
| NumberOfSentence | 0.054 | 0.076 |
| TextualSimilarity | 0.005 | 0.024 |
| *Codelength* | 0.234 | 0.153 |
| TFAnswerCode | 0.064 | 0.043 |
| *TFAnswerText* | 0.124 | 0.133 |
| Note: *Italics* denote noteworthy features | | |

## 6. DISCUSSION AND IMPLICATIONS

***RQ***. *Which features are most significant in distinguishing an accepted Stack Overflow answer?* Reflecting on our outcomes, both of our models show that the time lag between when a question was posted and an accepted answer was provided, the reputation of the user (contributor), the code length in posts and the textual similarity between the question and answer pairs were consistently the most dominant features that distinguished chosen Stack Overflow answers.

Our outcome here that **accepted answers are delayed** is surprising. Evidence has shown that Stack Overflow answers are typically accepted within a day, even though questions may start receiving answers in around 21 minutes [3]. This evidence suggests that those posting questions deliberate on the answers that are provided by the community before finally selecting an answer as accepted. In fact, our outcomes in this work suggest that these **members may have no choice but to wait for appropriate answers, which tend to take some time before they are logged by knowledgeable community members**. This evidence suggests that a utility to predict the next best (or acceptable) answer would be noteworthy, and particularly in instances where votes are not available with which to offer preliminary judgement on answers. In fact, our outcomes show that **the votes (Score) that Stack Overflow answers attracted did not influence their acceptability**.

Other significant features revealed in this work may provide insights to those posing questions in informing their selection of a most suitable answer. Our evidence also points to a potential lack of timeliness of acceptable answers being provided by the community, which could have negative consequences on software development time. While previous work has established that the questions that are asked on Stack Overflow usually receive one or more answers [1], it is preferable that acceptable answers are provided in the shortest possible time. This is necessary as software developers frequently consult Stack Overflow for solutions to their programming challenges, to the extent that Stack Overflow is becoming a substitute for official programming languages' tutorials [2]. If developers ask questions and must then wait for extended periods of times before acceptable answers are provided, they may lose interest in the Stack Overflow community.

Our outcomes show that **the reputation of the user was one of the most dominating features that distinguished a chosen Stack Overflow answer**. This evidence converges with the outcomes of previous work, which found users' reputation to be the strongest predictor of the quality of posts [12]. In the Stack Overflow community, members enhance their reputation through the votes they receive from their questions and answers. If a question or answer logged by a user is voted up that user receives 5 or 10 points respectively. Accepting an answer gives the acceptor 2 points, and the contributor providing the answer is awarded 15 points. Other means of enhancing reputation include through edits (2 points), upvotes (5 points) and changes approved (2 points). Contributors also lose reputation when their contributions are voted down (-1 point) or when posts are tagged as offensive or spam (-100 points). Through this complex mechanism of rewards, those contributors that are ranked more highly (or acquire more points) tend to stand out in terms of Stack Overflow answer acceptance. While on the one hand highly ranked members may indeed provide the best answers [5], on the other, these members' prestige may also enhance others' trust for their contributions. This could be problematic for the Stack Overflow community. For instance, the need to 'win' reputation rewards can at times influence answer quality, as users may be driven to provide an answer without considering the quality [6]. Contributors may also set out to game the Stack Overflow platform, by answering questions relating to popular or easy topics, which may in turn lead to an increase in their reputation. To this end, while community members may regard these contributors on the basis of the points they have acquired, such members may in fact not be as knowledgeable of certain topics as their reputation suggests. Thus, members' reputation should be considered in relation to other aspects of their answers (e.g., code length, considered below).

Evidence in this work shows that **code length in posts distinguished acceptable Stack Overflow answers**. Generally, those having more to say tend to write more, which in turn may be linked to their more knowledgeable demeanor. However, evidence of more code statements is not always linked to the quality of the solution that is provided. Previous studies have challenged this view; for instance, Jeon, et al. [6] established that the quality of an answer is related to its length. More specifically, previous work on code readability concluded that lines of code and the average number of identifiers per line predict the level of readability evident in code [16]. While it is not plausible that code length on its own is likely to predict answer acceptability, and in fact, evidence has shown that isolated predictors may be significant by chance [13]. , when combined, multiple variables observed in this work do interact to reflect Stack Overflow answer acceptability.

Our results show that the **textual similarity between the question and answer pairs also enhanced answers' acceptability**. Those answers that possessed syntactic relations with associated questions were more likely to be accepted. The work of Blooma, et al. [14] pointed out that the accuracy of an answer was the most significant contributing factor that determines the best answer. This was assessed based on how close an answer was to a given question in vector space. Our outcome in this work provides confirmation for this assessment, where we observed this feature to be among the top four established for distinguishing a chosen Stack Overflow answer. We anticipate that acceptable answers need to be contextualized, such that aspects of the answer should be situated in relation to request(s) in the question. This way, those reading the answer are able to make sense of the solution. This in turn likely leads to textual similarities, and thus, the evidence observed here.

Overall, the attributes (features) above were shown to be consistent predictors across two modelling approaches (random forest and neural network), adding credibility to our evaluations. The time lag between when a question was posted and an answer was provided, the reputation of the user (contributor), the code length in posts and the textual similarity between question and answer pairs were consistently the dominant features that distinguished chosen Stack Overflow answers. In fact, these features recorded similar predictive power in both models. To this end, our outcomes are somewhat divergent to those provided by the work of [14], which concluded that the best answers are greatly influenced by textual features only. In our case, code (e.g., code length), textual (e.g., textual similarity of question and answer), non-textual (e.g., time lag), and user (e.g., reputation of contributor) features were seen to have some level of influence on an answer being accepted (or not) on Stack Overflow. Moreover, these features may be strictly dichotomized across textual (e.g., textual similarity of question and answer) and non-textual (e.g., reputation of contributor) features, also diverging from earlier findings.

We contend that this outcome is of interest as we have examined the predictive power of a large number of features (textual and non-textual) in terms of discriminating accepted from unaccepted answers. To this end, we believe this study could be useful for software developers focused on developing plugins that integrate with IDEs in order to display Stack Overflow question and answer pairs. In fact, researchers and developers could use the results of our work to rank answers in their tool (based on the features that demonstrate significance). Also, in relation to software practitioners that use Stack Overflow for overcoming various challenges in their work (both in terms of developing high quality software and speed of development), we believe we have provided insights into the features to focus on when looking through answers that are not accepted or those that are low on community votes. For instance, under these circumstances, Stack Overflow administrators may implement a feature whereby answers are ordered based on the prediction of their acceptability. Our work also provides support for the *combination* of textual and non-textual attributes that influence answers that are accepted on Stack Overflow. That said, there is need for experiments to go one step further and to explore the effects of users changing their accepted answers over time, as this data was not examined in our study.

# 7. THREATS TO VALIDITY

Our dataset consisted of 249,588 records which were posted on Stack Overflow over three years (2014-2016). These may not represent all the types of questions asked and answers provided for posts tagged with Java or JavaScript on Stack Overflow. To this end, our results may not hold true for other categories of questions, although recent evidence suggests that there is consistency in trends across languages for aspects of Stack Overflow code [21]. That said, the Stack Overflow dataset is not representative of all Q&A portals, although we expect that portals dedicated to technology-related Q&A may hold similar content (e.g., Yahoo!Answers for programming). Thus, the features that discriminate accepted answers for such portals may be similar. Also, we have based the acceptable answer on the answer selected by the user who posed the question. This may not necessarily be the *best* answer in all cases, as some users may lack knowledge, affecting their ability to differentiate among answers. Furthermore, we have only used two modelling approaches in this work, and thus, we cannot definitively say that our outcomes will hold true for other modelling methods (e.g., Bayesian logistic regression or support vector machines (SVM)). Finally, we made some assumptions in Section 3 in discarding some of the features (i.e., badges and post history) in the dataset. These were discarded because we believe such features do not relate to the answer contained in the text. This issue may be validated through inductive analysis. These potential threats should be taken into account when evaluating our outcomes.

# 8. CONCLUSION AND FUTURE WORK

Mechanisms to determine the features that differentiate an acceptable answer from an unacceptable answer in a Q&A forum such as Stack Overflow could be used to build models that are able to rank the best answers available. Having such a ranked list of answers may be used to re-order appropriate answers, potentially making it easier for users of such a forum to more easily find solutions to their problems. In addition, developers of plugins and tools could create artifacts that are able to rank answers from such outcomes, and users of such plugins may then drag and drop such documents (Q&A pairs) into their programming environment.

The main objective of this paper was thus to explore the predictive power of textual and non-textual features in discriminating acceptable Stack Overflow answers. In this regard, we observed that the length of code, time lag, user reputation, and similarity of the text between questions and answers were features that characterized acceptable answers. We have provided insights into the key attributes for users to look for in distinguishing what makes a Stack Overflow answer acceptable. This outcome could be of practical significance to the many practitioners that use Stack Overflow to answer their questions, or developers interested in developing plugins which use the Stack Overflow dataset in reducing development time (in terms of ranking returned outcomes). Developers of such plugins could incorporate the features found to rank answers. This could be particularly valuable in instances where no answer is accepted on Stack Overflow, as users could consider those features revealed in this work in selecting the best possible answer. However, we concede that there is a need for follow-up work before we could fully encourage implementation of these recommendations on Stack Overflow and other community forums. Beyond Stack Overflow, we believe that the outcomes of this work may also aid Q&A feature selection more generally. Accordingly, a plausible next step for future work is to replicate this study for similar Q&A forums. We also plan to evaluate other feature selection algorithms and perform inductive analysis to validate our outcomes.